\documentclass[prl,twocolumn,graphicx,amssymb,floatfix]{revtex4}

\begin{document}

{\bf Comment on ``Weak value amplification is suboptimal for estimation and detection'' }

In a recent Letter,  Ferrie and  Combes \cite{FC} defined the practical tasks ``detect'' and ``estimate'' and    concluded that
``Post-selection cannot aid in detect and estimate for any interaction parameter''. In particular, they argued that
``there is no sense in which  WVA [Weak Value Amplification] provides an ``amplification'' for quantum metrology''.

At 1988 Aharonov, Albert and Vaidman \cite{AAV} discovered that a sufficiently weak coupling to any observable of a  pre- and postselected quantum system is a coupling to the ``weak value'' of this observable and, since the weak value can be much larger than the eigenvalues of the observable, this method provides an effective amplification of the weak coupling. This amplification is the WVA discussed in the Letter of Ferrie and  Combes. The WVA method has been implemented in several experiments in recent years. The spin Hall effect for light was first detected using this method \cite{HK}. A record precision of a mirror angle estimation was obtained using WVA in another experiment \cite{How}. So, definitely, the post-selection aided in detecting and estimating interaction parameters.

How can it then be that the ``statistically rigorous arguments'' of  Ferrie and  Combes contradict these experimental results? The explanation is that the assumptions in their statistical analysis are irrelevant for realistic experimental situations. 
I found the main erroneous  assumption which led Ferrie and Combes to their incorrect conclusions  thank to my  direct involvement in two weak measurement experiments \cite{Xi,Danan}. The limiting factor in these and other experiments is not the number of preselected quantum systems  (photons) considered by Ferrie and Combes, but the number of detected, post-selected photons. The saturation of the detectors generally happens much before the power limitation of the laser source kicks in. Thus, the low probability of the postseletction, the main negative factor in experiments with large weak values, is not relevant. This then undermines the conclusions of Ferrie and Combes.

In their Letter, Ferrie and Combes quote  other recent papers analyzing the limitations of the WVA method \cite{KnBr,TaYa,Zhu,KnGa}, which they improve and complement.
 These limitations  were  obtained by using the same assumptions, but the authors of these works specify (some of them maybe not clearly enough) that their conclusions are conditioned on these assumptions. Zhu et al. \cite{Zhu} do it very precisely. They conclude: ``We have shown that weak measurements cannot effectively improve
the SNR [Signal to Noise Ratio] and the  MS [measurement sensitivity] when the probability decrease due
to postselection needs to be considered; while for practical cases when the probability reduced by postselection need not
be considered, weak measurements can significantly improve both the SNR and the MS.''

This work has been supported in part by grant number 32/08 of the Binational Science Foundation and the Israel Science Foundation  Grant No. 1125/10.

L. Vaidman\\
 Raymond and Beverly Sackler School of Physics and Astronomy\\
 Tel-Aviv University, Tel-Aviv 69978, Israel

\end{document}